\documentclass[twocolumn,showpacs,preprintnumbers,amsmath,amssymb]{revtex4}

\usepackage{graphicx}
\usepackage{dcolumn}
\usepackage{bm}
\usepackage{epstopdf}
\usepackage{mathrsfs}
\usepackage{amssymb,amsmath,amsfonts,latexsym}
\usepackage{nicefrac}
\usepackage[colorlinks=true,linktocpage=true,linkcolor=blue,citecolor=blue]{hyperref}
\usepackage[utf8]{inputenc}
\usepackage{lipsum}
\usepackage{nicefrac}
\usepackage[normalem]{ulem}

\newcommand{\eqn}[1]{Eq.~\eqref{#1}}

\newcommand{\sst}{\scriptscriptstyle }
\long\def\comment#1{ }

\newcommand{\onefourth}{{\nicefrac{1}{4}}}
\newcommand{\onethird}{{\nicefrac{1}{3}}}

\def\0{{\boldsymbol 0}}

\def\Q{{\cal Q}}
\def\C{{\cal C}}
\def\P{{\cal P}}

\def\med{\text{med}}
\def\sing{\text{sing}}

\def\tform{{t_\text{f}}}
\def\tdecoh{t_\text{d}}

\def\pT{p_{\sst T}}

\def\pT{p_{_T}}

\newcommand{\beq}{\begin{eqnarray}}
\newcommand{\eeq}{\end{eqnarray}}
\newcommand{\be}{\begin{eqnarray*}}
\newcommand{\ee}{\end{eqnarray*}}
\newcommand{\bal}{\begin{align}}
\newcommand{\eal}{\end{align}}
\newcommand{\rmd}{{\rm d}}
\newcommand{\dd}{{\rm d}}
\newcommand{\rme}{{\rm e}}

\def\abar{{\rm \bar\alpha}}

\newcommand{\nn}{\nonumber\\ }

\begin{document}


\title{Sudakov suppression of jets in QCD media}

\author{Yacine Mehtar-Tani} 
\email{ymehtar@uw.edu}
\affiliation{Institute for Nuclear Theory, University of Washington, Box 351550, Seattle, WA 98195-1550, USA}
\author{Konrad Tywoniuk}
\email{konrad.tywoniuk@cern.ch}
\affiliation{Theoretical Physics Department, CERN, 1211 Geneva 23, Switzerland}%

\date{\today}

\begin{abstract}
We  compute modifications to the jet spectrum in the presence of a dense medium. We show that in the large-$N_c$ approximation and at leading logarithmic accuracy the jet nuclear modification factor factorizes into a quenching factor associated to the total jet color charge and  a Sudakov suppression factor which accounts for the energy loss of jet substructure fluctuations. This factor, called the jet collimator, implements the fact that subjets, that are not resolved by the medium, lose energy coherently as a single color charge, whereas resolved large angle fluctuations suffer more quenching. For comparison, we show that neglecting color coherence results in a stronger suppression of the jet nuclear modification factor. 
\end{abstract}

\preprint{INT-PUB-17-025, CERN-TH-2017-161}
\maketitle

The properties of fully reconstructed jets in heavy-ion collisions \cite{Aad:2010bu,Chatrchyan:2012nia,Aad:2014bxa,Abelev:2013kqa} reveal the effects of notable final state interactions. They are currently actively investigated as probes of the underlying deconfined hot matter produced in these collisions. A remarkable observation is the strong suppression of the jet yield that persists over a large range of transverse momentum. In contrast, in-cone jet modifications tend to decrease in the same variable \cite{Chatrchyan:2014ava,Aaboud:2017bzv,CMS:2016jys}. This challenges our understanding of the mechanisms underlying jet modifications in the presence of a QCD medium. 

In this context, quenching typically refers to the energy loss of jets caused by elastic and inelastic interactions with the medium.
The amount of energy, $\epsilon$, transported out of the jet cone is sampled by a probability distribution $\P(\epsilon)$, called the quenching weight. Hence, the final-state spectrum reads
\beq
\label{eq:spect-general}
\frac{\rmd \sigma_\med }{\rmd \pT^2\rmd y} = \int^\infty_0 \rmd \epsilon\, \P(\epsilon) \, \frac{\rmd \sigma_\text{vac} (\pT+\epsilon)}{\rmd\pT^2 \rmd y} \,,
\eeq
where $\rmd \sigma_\text{vac} $ is the jet spectrum in vacuum. 
The quenching probability distribution is expected to depend on the medium properties, such as the jet quenching parameter $\hat q$, which is a medium diffusion coefficient in transverse momentum space,  and the medium length $L$ \cite{Burke:2013yra}, but it should also be sensitive to the jet $\pT$ and cone size $R$. 

The tools for analyzing jet quenching were developed to account for radiative and elastic energy loss of a single color charge propagating in the medium \cite{Wang:1991xy,Baier:1996sk,Zakharov:1997uu,Wiedemann:2000za,Gyulassy:2000fs,Arnold:2002ja}, see also  recent reviews \cite{Mehtar-Tani:2013pia,Blaizot:2015lma} and references therein. 
However, owing to the QCD mass singularity, the jet-initiating parton tends to branch rapidly---in particular inside the medium. This leads one to question the single charge energy loss approximation for jet quenching at high-$\pT$. Radiative energy loss of color-connected subjets was recently shown to be sensitive to interference effects between the emitters \cite{Mehtar-Tani:2017ypq}, see also \cite{MehtarTani:2010ma,MehtarTani:2011tz,CasalderreySolana:2011rz,CasalderreySolana:2012ef}. On the other hand, most Monte Carlo implementations of jet quenching ignore interference effects and assume independent energy loss of all jet constituents. One may expect a substantial quantitative discrepancy between the two pictures. 

\begin{figure}
\includegraphics[width=0.95\columnwidth]{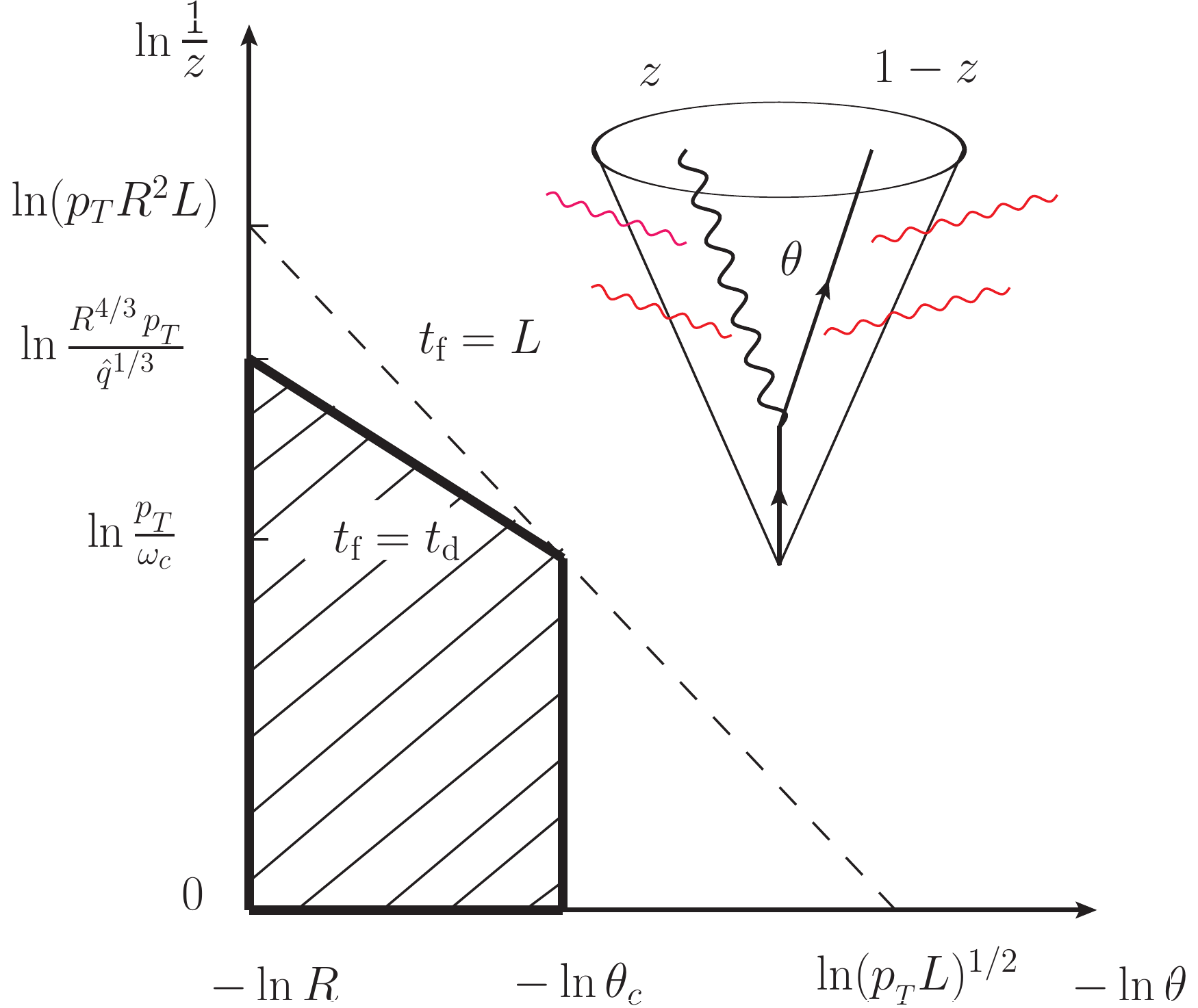}
\caption{The phase-space of higher-order corrections to the jet inclusive spectrum in the presence of radiative energy loss. We have incorporated lines denoting $\tform = L$, $\tform = \tdecoh$ and $\tdecoh = L$ (implying $\theta = \theta_c$) and all related scales.}
\label{fig:nloeloss}
\end{figure}

In vacuum \cite{Kinoshita:1962ur,Lee:1964is}, higher-order corrections to the fully inclusive jet spectrum are suppressed by powers of the coupling constant \footnote{Note that for small cone sizes, powers of $\alpha_s \ln R$ needs to be resummed.}.
In the medium however, induced energy loss of the final-state particles causes a mismatch between real and virtual diagrams that is accompanied by logarithms of the available phase space.
This takes place whenever the medium resolves the individual color charges created in hard splittings, namely at the time $t = \tdecoh \sim (\hat q  \,\theta^2)^{-1/3}$, when the transverse distance between the two daughters, $r_\perp\sim \theta \, t$, becomes of order the medium correlation length $(\hat q \, t)^{-1/2}$.  
This translates to a minimum characteristic angle $\theta_c \sim (\hat q \,L^3)^{-1/2}$ when $\tdecoh = L$, so that, for $R < \theta_c$, the jet is not resolved. In this case, the energy loss probability distribution is not sensitive to the fluctuations of the jet substructure.
An example of a higher-order correction and the related phase space are  depicted in Fig.~\ref{fig:nloeloss}. In the inset the thick lines represent hard, vacuum-like real emission while the thin lines represent multiple medium-induced emissions. The formation time of splitting is given by $\tform = 2 \pT/m^2$, where $m^2 = z(1-z)\pT^2 \theta^2$ is the invariant mass-squared of two-parton system.
The real contribution will be affected by the combined quark and gluon energy loss while for the virtual contribution only the quark quenching plays a role.
The mismatch appears when $\tform \sim (z\pT\theta^2)^{-1} \ll \tdecoh \ll L$, corresponding to vacuum splittings inside the medium. As a result the double logarithmic phase space for this correction, given by the area of the shaded area in Fig.~\ref{fig:nloeloss}, is enhanced by the jet scales
\beq
\label{eq:PhaseSpace}
\abar \int\limits_{\tform < \tdecoh<L}\frac{\rmd \theta}{\theta} \int \frac{\rmd z}{z} \equiv \abar \ln \frac{R}{\theta_c } \left (\ln \frac{\pT}{\omega_c}  + \frac{2}{3} \ln \frac{R}{\theta_c } \right).
\eeq
This calls for the resummation of higher-order contributions. 

The main purpose of this Letter is to compute the effect of fluctuations of the jet substructure on the jet spectrum \eqn{eq:spect-general}. As will become clear in the moment, it is convenient to directly consider the ratio of the jet spectrum in medium and the unmodified vacuum spectrum, known as the nuclear modification factor,
\beq
\label{eq:raa_def}
R_\text{jet} = \left(\frac{\rmd \sigma_\med }{\rmd \pT^2 \rmd y} \right) \Big/ \left( \frac{\rmd \sigma_\text{vac} }{\rmd \pT^2  \rmd y} \right) \,.
\eeq
If one assumes a steeply falling power spectrum with a constant power $n$, one can make the following approximation \cite{Baier:2001yt}
\beq 
\label{eq:exp-spect}
\frac{\rmd \sigma_\text{vac} (\pT+\epsilon)}{\rmd \pT^2 \rmd y} \approx \frac{\rmd \sigma_\text{vac}}{\rmd \pT^2 \rmd y} \,  \exp \left(-\frac{n \epsilon}{p_{_T}} \right)\,,
\eeq
which holds for small $\epsilon/\pT$ and large $n$. Corrections to the exponential spectrum for finite or momentum dependent index $n$ can be systematically calculated. Using \eqn{eq:exp-spect}, it becomes clear that the ratio (\ref{eq:raa_def}) is related to the Laplace transform of the quenching weight, $\tilde \P (\nu) = \int_0^\infty\dd \epsilon \,\P(\epsilon) \,\rme^{-\nu \epsilon}$, as follows
\beq
R_\text{jet} =\Q(\pT) \,,
\eeq
where $\Q(\pT) \equiv \tilde \P\left(n\big/\pT\right)$.

Let us presently describe the effects of the fluctuating substructure on the total energy loss that a jet experiences, determining it from a perturbative expansion in what follows.
At leading order in the strong coupling constant, $\Q(\pT) $ is related to the quenching of a single quark \cite{Baier:2001yt}, $\Q^{(0)}(\pT) \equiv \Q_q(\pT)$. In the approximation of independent soft medium-induced radiation the quenching factor, $\Q_q(\pT) = \tilde \P_q(n/\pT)$, reads
\beq
\label{eq:lo-quenching}
\Q_q(\pT) = \exp\left[\int_0^\infty \rmd \omega \, \frac{\rmd I}{\rmd \omega}\,(\rme^{- n \omega / \pT}-1)\right] \,,
\eeq
where the medium induced radiation spectrum is given by $\omega \rmd I/\rmd \omega =2\abar\ln|\cos(1+i)\sqrt{\omega_c /2\omega} |$ \cite{Baier:1996sk,Zakharov:1997uu}, $\omega_c \equiv \hat q L^2/2$.  In \eqn{eq:lo-quenching}, the coupling constant $\bar \alpha \equiv \alpha_s C_{\sst F}/\pi$ should be evaluated at the transverse scale $k_\text{f}(\omega) \equiv \sqrt{\hat q \tform}= (2 \hat q \omega)^{1/4}$ of the radiated gluons, but since the integral in \eqn{eq:lo-quenching} is dominated by $\omega \sim \pT/n$,  to logarithmic accuracy, it can be taken to run as $\alpha_s\big((\pT\hat q/n)^{1/4}\big)$.

When $\omega\sim  \pT/n \ll \omega_c$, the radiation spectrum can be approximated by $\abar\sqrt{\omega_c/2\omega}$, which yields the quenching factor $\Q_q(\pT)  = \exp\left(-\sqrt{2\pi n\abar^2 \omega_c /\pT }\right)$ \cite{Baier:2001yt}.  But in general, the quenching factor depends on the jet radius. In the regime of interest, $Q_q \ll 1$ which corresponds to gluon frequencies, $\pT/ n\lesssim \abar^2 \omega_c$, the radiated gluons undergo a democratic cascading process and eventually, lose all of their energy to the medium \cite{Blaizot:2013hx}. Therefore, their energy is uniformly distributed in angles and for small jet radii their contribution inside the jet cone can be neglected. On the other hand, small angle hard medium induced radiation $\omega \gtrsim \bar \alpha^2\omega_c$ are rare, i.e. $\mathcal{O}(\alpha_s)$ corrections, and thus can be treated as higher order contributions. The characteristic angle that separates the two regimes is $\theta_\text{f}(\abar^2\omega_c)\sim \abar^{-3/2}  \theta_c \gg \theta_c$, where we have used that $\theta_\text{f}(\omega)\equiv k_\text{f}(\omega) /\omega=(\hat q/\omega^3)^{1/4}$. Hence, in what follows we shall work in the approximation $R \ll \abar^{-3/2} \theta_c $. This allows for an angular separation between medium-induced gluon radiation and collinear enhanced vacuum splittings. 

Throughout, we will work in the large-$N_c$ limit and consider only soft medium-induced gluons. We will also assume the dominance of quark jets, described by a hard spectrum with constant spectral index $n$.

For high-$\pT$ jets, the logarithmically enhanced contribution to the quenching at first order in the coupling constant, $\Q^{(1)}(\pT)$, arises in the region where the formation time of the hard quark-gluon pair is shorter than the medium length, i.e. when the splitting forms promptly in the medium and propagates through approximately the whole length of the medium. It reads
\beq
\!\Q^{(1)}( \pT) = \frac{\alpha_s}{\pi}\!\!\int_0^1 \!\rmd z  P_{gq}(z) \!\int_0^R \frac{\rmd \theta}{\theta} \Big[ \Q_{qg}( \pT)- \Q_q( \pT)\Big],
\label{eq:nlo-quenching}
\eeq
for fixed coupling,  where 
$P_{gq}(z)=C_F[1+(1-z)^2]/z$ is the quark-gluon Altarelli-Parisi splitting function, with $z$ the gluon momentum fraction, and $\Q_{qg}(\pT)=\tilde\P_{qg}\big(n/\pT,\theta\big)$  is the quenching factor of the promptly produced quark-gluon pair with opening angle $\theta$  \cite{Mehtar-Tani:2017ypq}. 
The above correction is negative since two prongs are expected to suffer more energy loss than one, i.e., $\Q_{qg} < \Q_q$. 
In this expression, we have assumed  $z \pT \gg \abar^2 \omega_c $, which leaves the argument of the splitting function unmodified. Indeed, when $z \pT <   \abar^2 \omega_c$, then the minimum splitting angle $\theta_\text{f}(z\pT)$ that follows from the constraint, $\tform \ll \tdecoh$, satisfies $\theta_\text{f}(z\pT)> \abar^{-3/2} \theta_c$, which is assumed to be parametrically larger than $R$. 

In the large-$N_c$ approximation, 
$\Q_{qg}(\pT) = \Q_q(\pT) \Q_\sing(\pT )$,
where $\Q_\sing(\pT) \equiv \tilde \P_\sing\big(n/\pT,\theta \big)$ is the quenching factor of a quark-antiquark singlet antenna.
The singlet quenching factor in the limit $\tform \ll \tdecoh$ (see also \cite{Casalderrey-Solana:2015bww}), which explicitly depends on the angle of the fluctuation, is found from 
$\Q_\sing(\pT)  = \Q_q^2(\pT,L) - 2 \int_0^L \!\!\dd t \,  S_2(t)\,\gamma(\pT)\Q_q^2\big(\pT,L-t \big),$
where we have restored the time-dependence in the single-parton quenching weights on the right-hand-side and 
$\gamma(\pT,t) = \int_0^\infty \rmd \omega\, \frac{\rmd I}{\rmd \omega\rmd t} \, (1-\rme^{-n \omega /\pT })$ \cite{Mehtar-Tani:2017ypq}.
For further details on the interference spectrum, see \cite{CasalderreySolana:2011rz,MehtarTani:2012cy}.
Finally, $S_2 (t)= \exp \left(- \hat q \,\theta^2 t^3/12 \right)$ describes the color decoherence of the pair that is sensitive to the characteristic time scale $\tdecoh \sim (\hat q \,\theta^2)^{-\onethird}$ \cite{MehtarTani:2011tz}. At later times, interferences are suppressed and the gluon and quark radiate independently.

For the purpose of this work, it is sufficient to identify two limiting cases for $\Q_\sing(\pT) $ in order to extract the regimes of logarithmic enhancement. When $\tdecoh \gg L$, which  translates into the small angle region $\theta \ll  \theta_c\equiv  (\hat q L^3/12)^{-1/2}$, we have $\Q_\sing(\pT)\simeq 1$ and when $\tdecoh \ll L$, i.e. at large angles $\theta \gg \theta_c$, we have $\Q_\sing(\pT) \simeq \Q^2_q(\pT)$. We note that in the former situation $\Q_{qg}\simeq \Q_q$ and, as a result, \eqn{eq:nlo-quenching} vanishes.  Hence, only the region $\theta >\theta_c$ contributes to logarithmic accuracy. This conveys the main physics message, namely that the medium resolves only sufficiently wide jet fluctuations. 
When $\tform > \tdecoh$, the medium resolves the quark-gluon fluctuation resulting in a hard medium-induced splitting that does not exhibit a double-logarithmic structure particular of vacuum splittings.  Finally, vacuum splittings outside the medium cancel out in \eqn{eq:nlo-quenching}.

Let us now introduce a new object, that we call the collimator function, which encodes the quenching of wide angle jet fluctuations. It is defined as,
\beq
\label{eq:coll-def}
\C(\pT)\equiv \frac{\Q(\pT)}{\Q_q(\pT)} \,,
\eeq
and represents the total quenching factor modulo the energy loss of the total color charge of the (quark) jet. The expansion of the collimator takes the form $\C(\pT) = 1+ \C^{(1)}(\pT) + \mathcal{O}(\alpha_s^2)$, where the first non-trivial correction appears at next-to-leading order.
We focus on the leading logarithmic (LL) behavior, that is $z \ll 1$ and $\theta \gg \theta_c$. 
In this situation, one can approximate $\Q_\sing(\pT) \simeq \Q^2_q(\pT)$,  
then \eqn{eq:coll-def} yields
\beq
\C^{(1)}(\pT) \simeq 2 \abar \int_{\theta_c}^R \frac{\rmd \theta}{\theta} \int_{(\hat q/\theta^4)^\onethird}^{\pT} \frac{\rmd \omega}{\omega}\,  \left[\Q^2_q(\pT) -1 \right] \,.
\label{eq:raa_nlo_log}
\eeq
Here we have denoted $\omega = z \pT$ and explicitly demanded that we must have $\tform < \tdecoh<L$
to preserve the collinear logarithm. This condition implies that $ \theta >  \theta_\text{f}(\omega)=(\hat q/ \omega^3)^\onefourth$ or equivalently  $ \omega > (\hat q /\,\theta^4)^{\onethird}$. 
Hence, the NLO result in the LL approximation becomes
\beq
\label{eq:raa_nlo_dlog}
\frac{\C^{(1)}(\pT)}{2 \abar \left[  \Q^2_{1}(p_{_T}) - 1\right]} \simeq \ln \frac{R}{\theta_c } \left (\ln \frac{\pT}{\omega_c}  + \frac{2}{3} \ln \frac{R}{\theta_c } \right) \,,
\eeq
for $\pT > \omega_c$, cf. Eq.~(\ref{eq:PhaseSpace}).
Remarkably, in the relevant high-$\pT$ regime, we obtain a single-logarithmic contribution that scales with the jet momentum.
In fact, this term arises from 
semi-hard radiation, $\pT > \omega > \omega_c$, and 
implies that the enhancement stems from splittings that appear early in the medium. 
A finite coherence angle $\theta_c$ moderates the enhancement, as 
will be demonstrated below.

The specific product of logarithms in \eqn{eq:raa_nlo_dlog} appears due to the fact that, at small angular scales, the jets are unresolved and lose energy coherently. Let us for the moment contrast this first-principle description \cite{Mehtar-Tani:2017ypq} with a scenario wherein we treat the jet substructures as completely independent with respect to medium interactions. The logarithmic phase space is only limited by the condition that the splittings happen inside the medium, i.e. $\tform < L$, cf. the Lund diagram in Fig.~\ref{fig:nloeloss}. Instead of the phase-space relevant for (\ref{eq:raa_nlo_log}), we now obtain a double-logarithmic enhancement,
\beq
\label{eq:DL-integral-decoherent}
\C^{(1)}(\pT) \Big|_\text{incoh} \simeq \frac{ \abar}{2}\left[\Q^2_{1}(\pT) -1 \right]\, \ln^2 \big( \pT R^2 L\big)  \,.
\eeq
This leads to a stronger energy-dependence and hence a stronger suppression than in \eqn{eq:raa_nlo_dlog}.

Let us now come back to dealing with the generalization of \eqn{eq:raa_nlo_dlog} to all orders. This is an arduous task which demands tracing multiple emissions with various formation times. In order to limit the scope, we will pursue the resummation only within the 
LL approximation. At this precision, all formation times can be assumed to be arbitrarily small as it corresponds to the phase-space giving rise to the maximal logarithmic enhancement. Writing the full dependence of the collimator function as $\C(\pT) \equiv \C(\pT,R)$ and assuming small energy losses, and hence neglecting any modification of the splitting functions, we can iterate the procedure by replacing
$\Q_q^2(\pT) \to \C_g (z,\pT,\theta) \C_q ((1-z),\pT,\theta)\,\Q_q^2(\pT) $, and 
$\Q_q(\pT) \to \C_q (1,\pT,\theta)\,\Q_q(\pT)$,
in the first and second terms of \eqn{eq:nlo-quenching}, respectively. This results in a non-linear, coupled evolution equation for the resummed collimator function for quarks and gluons,
\begin{align}
\label{eq:collimator-evol-2}
&\!\C_i (1,\pT,R)  =1 + \int_0^1 \rmd z  \int_{\theta_c}^R \frac{\rmd \theta}{\theta} \,\frac{\alpha_s(k_{\perp}) }{\pi}P_{gi}(z)\Theta(\tdecoh-\tform )
 \nn
&\!\!\times\!\!\Big[ \C_g (z,\pT,\theta) \C_i \big((1-z),\pT,\theta\big) \Q_q^2(\pT) - \C_i(1,\pT,\theta) \Big] \,,
\end{align}
where $i=q,g$ and we have restored the full Altarelli-Parisi splitting functions (cf. $P_{gq}(z) = C_F(1+(1-z)^2)^2/[z(1-z)]$). 
Here we have explicitly assigned the collimator functions to the quark and gluon daughters of the splitting.
Recall that it is implied that $\C (z,\pT,\theta) \equiv \C (z,\pT,\theta, L)$. Note also the explicit dependence on the energy fraction $z$ that is taken to be $1$ at the end of the evolution.
The restriction on the formation time implies that $(\hat q/\theta^4)^\onethird < z(1-z)E$. We have also introduced the one-loop running coupling 
$\alpha_s(k_{\perp})=1/(b\ln k_\perp^2/Q_0^2)$, where $k_{\perp} \equiv z(1-z) p_{\sst T} \theta$ and $b=(11-2n_f/3)/(4\pi)$, with $n_f = 5$ active flavors and $Q_0=250$ MeV. 
As usual, gluon splitting into quarks is neglected since it does not contribute at large-$N_c$ and does not contain IR divergences.
At logarithmic accuracy one has again used the fact that $ S_2$ 
provides a cut-off to the logarithmic angular integration $\theta > \theta_c$. 

Note that at very large jet transverse momenta, $\pT \gg n \abar^2 \hat q L^2$, $\Q_q \to 1$ and we see that $\C_q (\pT) =1 $  is indeed a fixed point of the equation. Hence, the expected large $\pT$ asymptotic limit is recovered. 
On the other hand, when $\Q_q(\pT) \ll 1$, for $\pT \ll n \abar^2 \hat q L^2 $,  one can neglect the  non-linear therm  in the evolution equation \eqn{eq:collimator-evol-2}.  At fixed coupling for instance, with  $\abar = \alpha_sC_F  /\pi$ for quarks, one obtains the exponentiation of \eqn{eq:raa_nlo_dlog} into the Sudakov form factor,
\beq
\label{eq:collimator-evol-final1}
\C_q(\pT,R)\simeq \exp\left[-2 \abar  \ln \frac{R}{\theta_c } \left (\ln \frac{p_{_T}}{\omega_c}  + \frac{2}{3} \ln \frac{R}{\theta_c } \right)\right],
\eeq 
when $\pT>\omega_c$, with the anticipated logarithmic phase space, and 
\beq
\label{eq:collimator-evol-final2}
\C_q(\pT,R)\simeq \exp\left[-\frac{3 \abar }{4} \ln^2 \frac{R^{4/3}p_{_T}}{\hat q ^{1/3}}  \right],
\eeq 
when $\pT<\omega_c$,
where we have only accounted for the soft limit of the splitting function. The gluon collimator in these limits follows from $\C_g = ( \C_q)^{N_c/C_F}$.

\begin{figure}[t!]
\centering
\includegraphics[width=\columnwidth]{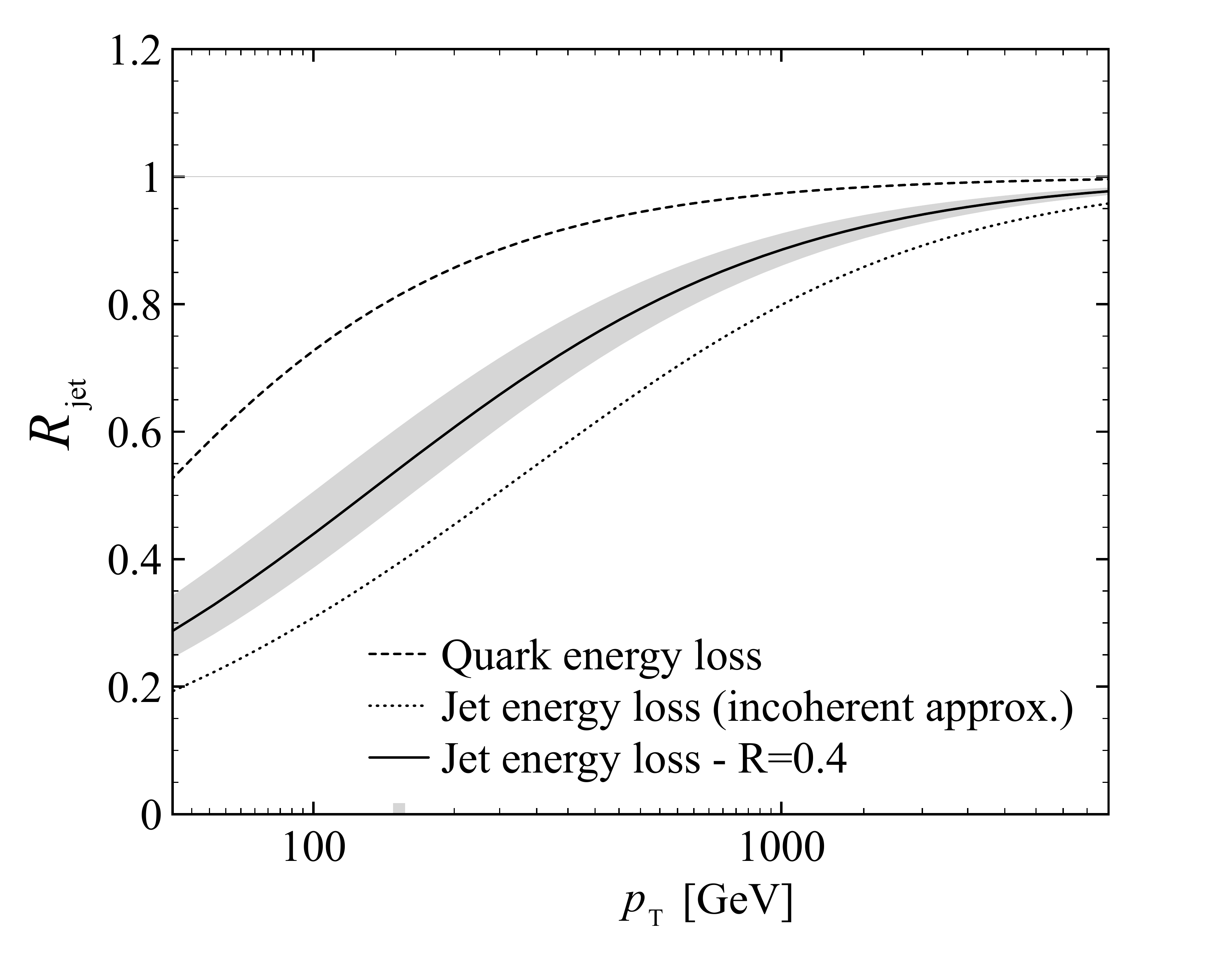}
\caption{The quark jet suppression factor calculated using the quenching weight alone (solid line) and additionally supplemented by the Sudakov suppression with and without coherence effects, see text for further details.}
\label{fig:jet-suppression}
\end{figure}
In order to have a quantitative estimate of the effect of the collimator function on the the nuclear modification factor and its potential in ruling out models that neglect color coherence, we have plotted the suppression factor for quark initiated jets,
\beq
\label{eq:rjet-final}
R_{\rm jet} = \Q_q(\pT) \times \C_q(\pT,R)\,,
\eeq
 in Fig.~\ref{fig:jet-suppression} compared to two limiting approximations. We have used index $n=6$ in the vacuum spectrum. 
The dashed line corresponds to the standard suppression arising from the energy loss of the total color charge, namely, $\C_q(\pT)=1$ in \eqn{eq:rjet-final}. The solid curve includes the additional Sudakov suppression due to jet collimation obtained by solving \eqn{eq:collimator-evol-2} for $R=0.4$  that accounts for coherence effects, while the band corresponds to a $\pm 0.1$ variation.
 The dotted line however does not account for coherence effects and exhibits stronger suppression as expected from the approximation leading to \eqn{eq:DL-integral-decoherent}. The chosen medium parameters were $\hat q = 1 $ GeV$^2$/fm and $L = 3$ fm. This corresponds to $\theta_c\simeq 0.13$ and $\omega_c\simeq 22.5$ GeV.

In conclusion, we have found that substructure fluctuations give rise to an additional suppression of the jet spectrum at high-$\pT$ in heavy-ion collisions beyond the energy loss experienced by the total color charge of the jet. At leading-logarithmic accuracy, the additional suppression originates from the quenching of resolved subjets that have been created early in the medium and hence propagate a large distance through it. We have resummed these contributions into a Sudakov suppression factor, called the collimator function since it suppresses large-angle jet fluctuations.
Coherence effects play an important role in moderating the effect of the collimator, giving rise to a single logarithm of the jet energy in contrast with  the double logs of the jet scale in the incoherent energy loss approximation. Our results demonstrate the sensitivity of inclusive jet observables to color coherence. It also opens the possibility to extend and refine studies of high-$\pT$ jet resummations in the presence of medium effects. We expect that the effects of fluctuating energy loss will have an impact on other jet quenching observables, see e.g. \cite{Milhano:2015mng,Casalderrey-Solana:2016jvj} for Monte Carlo studies.

Let us end with a final remark.  One generally expects to recover the missing energy by opening up the jet cone and hence an increase of the nuclear modification factor. This should be the result of the interplay between the suppression of high-$\pT$ jet fluctuations and the recapture of an increasing number of medium-induced soft particles within the jet cone. The physics of collimation produces an opposite trend, so for quantitative predictions it will also be important to implement the description of secondary medium-induced soft gluon emissions in order to improve the description of energy flow within and outside the jet cone.

The research of YMT is supported by the U.S. Department of Energy under Contract No. DE-FG02-00ER41132.
KT has been supported by a Marie Sklodowska-Curie Individual Fellowship of the European Commission's Horizon 2020 Programme under contract number 655279 ResolvedJetsHIC.

\bibliographystyle{apsrev4-1}
\bibliography{jetquenching}

\begin{thebibliography}{29}%
\makeatletter
\providecommand \@ifxundefined [1]{%
 \@ifx{#1\undefined}
}%
\providecommand \@ifnum [1]{%
 \ifnum #1\expandafter \@firstoftwo
 \else \expandafter \@secondoftwo
 \fi
}%
\providecommand \@ifx [1]{%
 \ifx #1\expandafter \@firstoftwo
 \else \expandafter \@secondoftwo
 \fi
}%
\providecommand \natexlab [1]{#1}%
\providecommand \enquote  [1]{``#1''}%
\providecommand \bibnamefont  [1]{#1}%
\providecommand \bibfnamefont [1]{#1}%
\providecommand \citenamefont [1]{#1}%
\providecommand \href@noop [0]{\@secondoftwo}%
\providecommand \href [0]{\begingroup \@sanitize@url \@href}%
\providecommand \@href[1]{\@@startlink{#1}\@@href}%
\providecommand \@@href[1]{\endgroup#1\@@endlink}%
\providecommand \@sanitize@url [0]{\catcode `\\12\catcode `\$12\catcode
  `\&12\catcode `\#12\catcode `\^12\catcode `\_12\catcode `\%12\relax}%
\providecommand \@@startlink[1]{}%
\providecommand \@@endlink[0]{}%
\providecommand \url  [0]{\begingroup\@sanitize@url \@url }%
\providecommand \@url [1]{\endgroup\@href {#1}{\urlprefix }}%
\providecommand \urlprefix  [0]{URL }%
\providecommand \Eprint [0]{\href }%
\providecommand \doibase [0]{http://dx.doi.org/}%
\providecommand \selectlanguage [0]{\@gobble}%
\providecommand \bibinfo  [0]{\@secondoftwo}%
\providecommand \bibfield  [0]{\@secondoftwo}%
\providecommand \translation [1]{[#1]}%
\providecommand \BibitemOpen [0]{}%
\providecommand \bibitemStop [0]{}%
\providecommand \bibitemNoStop [0]{.\EOS\space}%
\providecommand \EOS [0]{\spacefactor3000\relax}%
\providecommand \BibitemShut  [1]{\csname bibitem#1\endcsname}%
\let\auto@bib@innerbib\@empty
\bibitem [{\citenamefont {Aad}\ \emph {et~al.}(2010)\citenamefont {Aad} \emph
  {et~al.}}]{Aad:2010bu}%
  \BibitemOpen
  \bibfield  {author} {\bibinfo {author} {\bibfnamefont {G.}~\bibnamefont
  {Aad}} \emph {et~al.} (\bibinfo {collaboration} {ATLAS}),\ }\href {\doibase
  10.1103/PhysRevLett.105.252303} {\bibfield  {journal} {\bibinfo  {journal}
  {Phys. Rev. Lett.}\ }\textbf {\bibinfo {volume} {105}},\ \bibinfo {pages}
  {252303} (\bibinfo {year} {2010})},\ \Eprint {http://arxiv.org/abs/1011.6182}
  {arXiv:1011.6182 [hep-ex]} \BibitemShut {NoStop}%
\bibitem [{\citenamefont {Chatrchyan}\ \emph {et~al.}(2012)\citenamefont
  {Chatrchyan} \emph {et~al.}}]{Chatrchyan:2012nia}%
  \BibitemOpen
  \bibfield  {author} {\bibinfo {author} {\bibfnamefont {S.}~\bibnamefont
  {Chatrchyan}} \emph {et~al.} (\bibinfo {collaboration} {CMS}),\ }\href
  {\doibase 10.1016/j.physletb.2012.04.058} {\bibfield  {journal} {\bibinfo
  {journal} {Phys.Lett.}\ }\textbf {\bibinfo {volume} {B712}},\ \bibinfo
  {pages} {176} (\bibinfo {year} {2012})},\ \Eprint
  {http://arxiv.org/abs/1202.5022} {arXiv:1202.5022 [nucl-ex]} \BibitemShut
  {NoStop}%
\bibitem [{\citenamefont {Aad}\ \emph {et~al.}(2015)\citenamefont {Aad} \emph
  {et~al.}}]{Aad:2014bxa}%
  \BibitemOpen
  \bibfield  {author} {\bibinfo {author} {\bibfnamefont {G.}~\bibnamefont
  {Aad}} \emph {et~al.} (\bibinfo {collaboration} {ATLAS}),\ }\href {\doibase
  10.1103/PhysRevLett.114.072302} {\bibfield  {journal} {\bibinfo  {journal}
  {Phys. Rev. Lett.}\ }\textbf {\bibinfo {volume} {114}},\ \bibinfo {pages}
  {072302} (\bibinfo {year} {2015})},\ \Eprint {http://arxiv.org/abs/1411.2357}
  {arXiv:1411.2357 [hep-ex]} \BibitemShut {NoStop}%
\bibitem [{\citenamefont {Abelev}\ \emph {et~al.}(2014)\citenamefont {Abelev}
  \emph {et~al.}}]{Abelev:2013kqa}%
  \BibitemOpen
  \bibfield  {author} {\bibinfo {author} {\bibfnamefont {B.}~\bibnamefont
  {Abelev}} \emph {et~al.} (\bibinfo {collaboration} {ALICE}),\ }\href
  {\doibase 10.1007/JHEP03(2014)013} {\bibfield  {journal} {\bibinfo  {journal}
  {JHEP}\ }\textbf {\bibinfo {volume} {03}},\ \bibinfo {pages} {013} (\bibinfo
  {year} {2014})},\ \Eprint {http://arxiv.org/abs/1311.0633} {arXiv:1311.0633
  [nucl-ex]} \BibitemShut {NoStop}%
\bibitem [{\citenamefont {Chatrchyan}\ \emph {et~al.}(2014)\citenamefont
  {Chatrchyan} \emph {et~al.}}]{Chatrchyan:2014ava}%
  \BibitemOpen
  \bibfield  {author} {\bibinfo {author} {\bibfnamefont {S.}~\bibnamefont
  {Chatrchyan}} \emph {et~al.} (\bibinfo {collaboration} {CMS}),\ }\href
  {\doibase 10.1103/PhysRevC.90.024908} {\bibfield  {journal} {\bibinfo
  {journal} {Phys. Rev.}\ }\textbf {\bibinfo {volume} {C90}},\ \bibinfo {pages}
  {024908} (\bibinfo {year} {2014})},\ \Eprint {http://arxiv.org/abs/1406.0932}
  {arXiv:1406.0932 [nucl-ex]} \BibitemShut {NoStop}%
\bibitem [{\citenamefont {Aaboud}\ \emph {et~al.}(2017)\citenamefont {Aaboud}
  \emph {et~al.}}]{Aaboud:2017bzv}%
  \BibitemOpen
  \bibfield  {author} {\bibinfo {author} {\bibfnamefont {M.}~\bibnamefont
  {Aaboud}} \emph {et~al.} (\bibinfo {collaboration} {ATLAS}),\ }\href
  {\doibase 10.1140/epjc/s10052-017-4915-5} {\bibfield  {journal} {\bibinfo
  {journal} {Eur. Phys. J.}\ }\textbf {\bibinfo {volume} {C77}},\ \bibinfo
  {pages} {379} (\bibinfo {year} {2017})},\ \Eprint
  {http://arxiv.org/abs/1702.00674} {arXiv:1702.00674 [hep-ex]} \BibitemShut
  {NoStop}%
\bibitem [{\citenamefont {{CMS Collaboration}}(2016)}]{CMS:2016jys}%
  \BibitemOpen
  \bibfield  {author} {\bibinfo {author} {\bibnamefont {{CMS Collaboration}}},\
  }\href@noop {} {\  (\bibinfo {year} {2016})},\ \Eprint
  {http://arxiv.org/abs/CMS-PAS-HIN-16-006} {CMS-PAS-HIN-16-006} \BibitemShut
  {NoStop}%
\bibitem [{\citenamefont {Burke}\ \emph {et~al.}(2014)\citenamefont {Burke}
  \emph {et~al.}}]{Burke:2013yra}%
  \BibitemOpen
  \bibfield  {author} {\bibinfo {author} {\bibfnamefont {K.~M.}\ \bibnamefont
  {Burke}} \emph {et~al.} (\bibinfo {collaboration} {JET}),\ }\href {\doibase
  10.1103/PhysRevC.90.014909} {\bibfield  {journal} {\bibinfo  {journal} {Phys.
  Rev.}\ }\textbf {\bibinfo {volume} {C90}},\ \bibinfo {pages} {014909}
  (\bibinfo {year} {2014})},\ \Eprint {http://arxiv.org/abs/1312.5003}
  {arXiv:1312.5003 [nucl-th]} \BibitemShut {NoStop}%
\bibitem [{\citenamefont {Wang}\ and\ \citenamefont
  {Gyulassy}(1992)}]{Wang:1991xy}%
  \BibitemOpen
  \bibfield  {author} {\bibinfo {author} {\bibfnamefont {X.-N.}\ \bibnamefont
  {Wang}}\ and\ \bibinfo {author} {\bibfnamefont {M.}~\bibnamefont
  {Gyulassy}},\ }\href {\doibase 10.1103/PhysRevLett.68.1480} {\bibfield
  {journal} {\bibinfo  {journal} {Phys.Rev.Lett.}\ }\textbf {\bibinfo {volume}
  {68}},\ \bibinfo {pages} {1480} (\bibinfo {year} {1992})}\BibitemShut
  {NoStop}%
\bibitem [{\citenamefont {Baier}\ \emph {et~al.}(1997)\citenamefont {Baier},
  \citenamefont {Dokshitzer}, \citenamefont {Mueller}, \citenamefont {Peigne},\
  and\ \citenamefont {Schiff}}]{Baier:1996sk}%
  \BibitemOpen
  \bibfield  {author} {\bibinfo {author} {\bibfnamefont {R.}~\bibnamefont
  {Baier}}, \bibinfo {author} {\bibfnamefont {Y.~L.}\ \bibnamefont
  {Dokshitzer}}, \bibinfo {author} {\bibfnamefont {A.~H.}\ \bibnamefont
  {Mueller}}, \bibinfo {author} {\bibfnamefont {S.}~\bibnamefont {Peigne}}, \
  and\ \bibinfo {author} {\bibfnamefont {D.}~\bibnamefont {Schiff}},\ }\href
  {\doibase 10.1016/S0550-3213(96)00581-0} {\bibfield  {journal} {\bibinfo
  {journal} {Nucl.Phys.}\ }\textbf {\bibinfo {volume} {B484}},\ \bibinfo
  {pages} {265} (\bibinfo {year} {1997})}\BibitemShut {NoStop}%
\bibitem [{\citenamefont {Zakharov}(1997)}]{Zakharov:1997uu}%
  \BibitemOpen
  \bibfield  {author} {\bibinfo {author} {\bibfnamefont {B.}~\bibnamefont
  {Zakharov}},\ }\href {\doibase 10.1134/1.567389} {\bibfield  {journal}
  {\bibinfo  {journal} {JETP Lett.}\ }\textbf {\bibinfo {volume} {65}},\
  \bibinfo {pages} {615} (\bibinfo {year} {1997})}\BibitemShut {NoStop}%
\bibitem [{\citenamefont {Wiedemann}(2000)}]{Wiedemann:2000za}%
  \BibitemOpen
  \bibfield  {author} {\bibinfo {author} {\bibfnamefont {U.~A.}\ \bibnamefont
  {Wiedemann}},\ }\href {\doibase 10.1016/S0550-3213(00)00457-0} {\bibfield
  {journal} {\bibinfo  {journal} {Nucl. Phys.}\ }\textbf {\bibinfo {volume}
  {B588}},\ \bibinfo {pages} {303} (\bibinfo {year} {2000})},\ \Eprint
  {http://arxiv.org/abs/hep-ph/0005129} {arXiv:hep-ph/0005129 [hep-ph]}
  \BibitemShut {NoStop}%
\bibitem [{\citenamefont {Gyulassy}\ \emph {et~al.}(2000)\citenamefont
  {Gyulassy}, \citenamefont {Levai},\ and\ \citenamefont
  {Vitev}}]{Gyulassy:2000fs}%
  \BibitemOpen
  \bibfield  {author} {\bibinfo {author} {\bibfnamefont {M.}~\bibnamefont
  {Gyulassy}}, \bibinfo {author} {\bibfnamefont {P.}~\bibnamefont {Levai}}, \
  and\ \bibinfo {author} {\bibfnamefont {I.}~\bibnamefont {Vitev}},\ }\href
  {\doibase 10.1103/PhysRevLett.85.5535} {\bibfield  {journal} {\bibinfo
  {journal} {Phys. Rev. Lett.}\ }\textbf {\bibinfo {volume} {85}},\ \bibinfo
  {pages} {5535} (\bibinfo {year} {2000})},\ \Eprint
  {http://arxiv.org/abs/nucl-th/0005032} {arXiv:nucl-th/0005032 [nucl-th]}
  \BibitemShut {NoStop}%
\bibitem [{\citenamefont {Arnold}\ \emph {et~al.}(2002)\citenamefont {Arnold},
  \citenamefont {Moore},\ and\ \citenamefont {Yaffe}}]{Arnold:2002ja}%
  \BibitemOpen
  \bibfield  {author} {\bibinfo {author} {\bibfnamefont {P.~B.}\ \bibnamefont
  {Arnold}}, \bibinfo {author} {\bibfnamefont {G.~D.}\ \bibnamefont {Moore}}, \
  and\ \bibinfo {author} {\bibfnamefont {L.~G.}\ \bibnamefont {Yaffe}},\ }\href
  {\doibase 10.1088/1126-6708/2002/06/030} {\bibfield  {journal} {\bibinfo
  {journal} {JHEP}\ }\textbf {\bibinfo {volume} {06}},\ \bibinfo {pages} {030}
  (\bibinfo {year} {2002})},\ \Eprint {http://arxiv.org/abs/hep-ph/0204343}
  {arXiv:hep-ph/0204343 [hep-ph]} \BibitemShut {NoStop}%
\bibitem [{\citenamefont {Mehtar-Tani}\ \emph {et~al.}(2013)\citenamefont
  {Mehtar-Tani}, \citenamefont {Milhano},\ and\ \citenamefont
  {Tywoniuk}}]{Mehtar-Tani:2013pia}%
  \BibitemOpen
  \bibfield  {author} {\bibinfo {author} {\bibfnamefont {Y.}~\bibnamefont
  {Mehtar-Tani}}, \bibinfo {author} {\bibfnamefont {J.~G.}\ \bibnamefont
  {Milhano}}, \ and\ \bibinfo {author} {\bibfnamefont {K.}~\bibnamefont
  {Tywoniuk}},\ }\href {\doibase 10.1142/S0217751X13400137} {\bibfield
  {journal} {\bibinfo  {journal} {Int. J. Mod. Phys.}\ }\textbf {\bibinfo
  {volume} {A28}},\ \bibinfo {pages} {1340013} (\bibinfo {year} {2013})},\
  \Eprint {http://arxiv.org/abs/1302.2579} {arXiv:1302.2579 [hep-ph]}
  \BibitemShut {NoStop}%
\bibitem [{\citenamefont {Blaizot}\ and\ \citenamefont
  {Mehtar-Tani}(2015)}]{Blaizot:2015lma}%
  \BibitemOpen
  \bibfield  {author} {\bibinfo {author} {\bibfnamefont {J.-P.}\ \bibnamefont
  {Blaizot}}\ and\ \bibinfo {author} {\bibfnamefont {Y.}~\bibnamefont
  {Mehtar-Tani}},\ }\href {\doibase 10.1142/S021830131530012X} {\bibfield
  {journal} {\bibinfo  {journal} {Int. J. Mod. Phys.}\ }\textbf {\bibinfo
  {volume} {E24}},\ \bibinfo {pages} {1530012} (\bibinfo {year} {2015})},\
  \Eprint {http://arxiv.org/abs/1503.05958} {arXiv:1503.05958 [hep-ph]}
  \BibitemShut {NoStop}%
\bibitem [{\citenamefont {Mehtar-Tani}\ and\ \citenamefont
  {Tywoniuk}(2017)}]{Mehtar-Tani:2017ypq}%
  \BibitemOpen
  \bibfield  {author} {\bibinfo {author} {\bibfnamefont {Y.}~\bibnamefont
  {Mehtar-Tani}}\ and\ \bibinfo {author} {\bibfnamefont {K.}~\bibnamefont
  {Tywoniuk}},\ }\href@noop {} {\  (\bibinfo {year} {2017})},\ \Eprint
  {http://arxiv.org/abs/1706.06047} {arXiv:1706.06047 [hep-ph]} \BibitemShut
  {NoStop}%
\bibitem [{\citenamefont {Mehtar-Tani}\ \emph {et~al.}(2011)\citenamefont
  {Mehtar-Tani}, \citenamefont {Salgado},\ and\ \citenamefont
  {Tywoniuk}}]{MehtarTani:2010ma}%
  \BibitemOpen
  \bibfield  {author} {\bibinfo {author} {\bibfnamefont {Y.}~\bibnamefont
  {Mehtar-Tani}}, \bibinfo {author} {\bibfnamefont {C.~A.}\ \bibnamefont
  {Salgado}}, \ and\ \bibinfo {author} {\bibfnamefont {K.}~\bibnamefont
  {Tywoniuk}},\ }\href {\doibase 10.1103/PhysRevLett.106.122002} {\bibfield
  {journal} {\bibinfo  {journal} {Phys.Rev.Lett.}\ }\textbf {\bibinfo {volume}
  {106}},\ \bibinfo {pages} {122002} (\bibinfo {year} {2011})}\BibitemShut
  {NoStop}%
\bibitem [{\citenamefont {Mehtar-Tani}\ \emph
  {et~al.}(2012{\natexlab{a}})\citenamefont {Mehtar-Tani}, \citenamefont
  {Salgado},\ and\ \citenamefont {Tywoniuk}}]{MehtarTani:2011tz}%
  \BibitemOpen
  \bibfield  {author} {\bibinfo {author} {\bibfnamefont {Y.}~\bibnamefont
  {Mehtar-Tani}}, \bibinfo {author} {\bibfnamefont {C.~A.}\ \bibnamefont
  {Salgado}}, \ and\ \bibinfo {author} {\bibfnamefont {K.}~\bibnamefont
  {Tywoniuk}},\ }\href {\doibase 10.1016/j.physletb.2011.12.042} {\bibfield
  {journal} {\bibinfo  {journal} {Phys.Lett.}\ }\textbf {\bibinfo {volume}
  {B707}},\ \bibinfo {pages} {156} (\bibinfo {year}
  {2012}{\natexlab{a}})}\BibitemShut {NoStop}%
\bibitem [{\citenamefont {Casalderrey-Solana}\ and\ \citenamefont
  {Iancu}(2011)}]{CasalderreySolana:2011rz}%
  \BibitemOpen
  \bibfield  {author} {\bibinfo {author} {\bibfnamefont {J.}~\bibnamefont
  {Casalderrey-Solana}}\ and\ \bibinfo {author} {\bibfnamefont
  {E.}~\bibnamefont {Iancu}},\ }\href {\doibase 10.1007/JHEP08(2011)015}
  {\bibfield  {journal} {\bibinfo  {journal} {JHEP}\ }\textbf {\bibinfo
  {volume} {1108}},\ \bibinfo {pages} {015} (\bibinfo {year}
  {2011})}\BibitemShut {NoStop}%
\bibitem [{\citenamefont {Casalderrey-Solana}\ \emph
  {et~al.}(2013)\citenamefont {Casalderrey-Solana}, \citenamefont
  {Mehtar-Tani}, \citenamefont {Salgado},\ and\ \citenamefont
  {Tywoniuk}}]{CasalderreySolana:2012ef}%
  \BibitemOpen
  \bibfield  {author} {\bibinfo {author} {\bibfnamefont {J.}~\bibnamefont
  {Casalderrey-Solana}}, \bibinfo {author} {\bibfnamefont {Y.}~\bibnamefont
  {Mehtar-Tani}}, \bibinfo {author} {\bibfnamefont {C.~A.}\ \bibnamefont
  {Salgado}}, \ and\ \bibinfo {author} {\bibfnamefont {K.}~\bibnamefont
  {Tywoniuk}},\ }\href {\doibase 10.1016/j.physletb.2013.07.046} {\bibfield
  {journal} {\bibinfo  {journal} {Phys. Lett.}\ }\textbf {\bibinfo {volume}
  {B725}},\ \bibinfo {pages} {357} (\bibinfo {year} {2013})},\ \Eprint
  {http://arxiv.org/abs/1210.7765} {arXiv:1210.7765 [hep-ph]} \BibitemShut
  {NoStop}%
\bibitem [{\citenamefont {Kinoshita}(1962)}]{Kinoshita:1962ur}%
  \BibitemOpen
  \bibfield  {author} {\bibinfo {author} {\bibfnamefont {T.}~\bibnamefont
  {Kinoshita}},\ }\href {\doibase 10.1063/1.1724268} {\bibfield  {journal}
  {\bibinfo  {journal} {J. Math. Phys.}\ }\textbf {\bibinfo {volume} {3}},\
  \bibinfo {pages} {650} (\bibinfo {year} {1962})}\BibitemShut {NoStop}%
\bibitem [{\citenamefont {Lee}\ and\ \citenamefont
  {Nauenberg}(1964)}]{Lee:1964is}%
  \BibitemOpen
  \bibfield  {author} {\bibinfo {author} {\bibfnamefont {T.~D.}\ \bibnamefont
  {Lee}}\ and\ \bibinfo {author} {\bibfnamefont {M.}~\bibnamefont
  {Nauenberg}},\ }\href {\doibase 10.1103/PhysRev.133.B1549} {\bibfield
  {journal} {\bibinfo  {journal} {Phys. Rev.}\ }\textbf {\bibinfo {volume}
  {133}},\ \bibinfo {pages} {B1549} (\bibinfo {year} {1964})}\BibitemShut
  {NoStop}%
\bibitem [{\citenamefont {Baier}\ \emph {et~al.}(2001)\citenamefont {Baier},
  \citenamefont {Dokshitzer}, \citenamefont {Mueller},\ and\ \citenamefont
  {Schiff}}]{Baier:2001yt}%
  \BibitemOpen
  \bibfield  {author} {\bibinfo {author} {\bibfnamefont {R.}~\bibnamefont
  {Baier}}, \bibinfo {author} {\bibfnamefont {Y.~L.}\ \bibnamefont
  {Dokshitzer}}, \bibinfo {author} {\bibfnamefont {A.~H.}\ \bibnamefont
  {Mueller}}, \ and\ \bibinfo {author} {\bibfnamefont {D.}~\bibnamefont
  {Schiff}},\ }\href@noop {} {\bibfield  {journal} {\bibinfo  {journal} {JHEP}\
  }\textbf {\bibinfo {volume} {0109}},\ \bibinfo {pages} {033} (\bibinfo {year}
  {2001})},\ \Eprint {http://arxiv.org/abs/hep-ph/0106347}
  {arXiv:hep-ph/0106347 [hep-ph]} \BibitemShut {NoStop}%
\bibitem [{\citenamefont {Blaizot}\ \emph {et~al.}(2013)\citenamefont
  {Blaizot}, \citenamefont {Iancu},\ and\ \citenamefont
  {Mehtar-Tani}}]{Blaizot:2013hx}%
  \BibitemOpen
  \bibfield  {author} {\bibinfo {author} {\bibfnamefont {J.-P.}\ \bibnamefont
  {Blaizot}}, \bibinfo {author} {\bibfnamefont {E.}~\bibnamefont {Iancu}}, \
  and\ \bibinfo {author} {\bibfnamefont {Y.}~\bibnamefont {Mehtar-Tani}},\
  }\href {\doibase 10.1103/PhysRevLett.111.052001} {\bibfield  {journal}
  {\bibinfo  {journal} {Phys. Rev. Lett.}\ }\textbf {\bibinfo {volume} {111}},\
  \bibinfo {pages} {052001} (\bibinfo {year} {2013})},\ \Eprint
  {http://arxiv.org/abs/1301.6102} {arXiv:1301.6102 [hep-ph]} \BibitemShut
  {NoStop}%
\bibitem [{\citenamefont {Casalderrey-Solana}\ \emph
  {et~al.}(2016)\citenamefont {Casalderrey-Solana}, \citenamefont {Pablos},\
  and\ \citenamefont {Tywoniuk}}]{Casalderrey-Solana:2015bww}%
  \BibitemOpen
  \bibfield  {author} {\bibinfo {author} {\bibfnamefont {J.}~\bibnamefont
  {Casalderrey-Solana}}, \bibinfo {author} {\bibfnamefont {D.}~\bibnamefont
  {Pablos}}, \ and\ \bibinfo {author} {\bibfnamefont {K.}~\bibnamefont
  {Tywoniuk}},\ }\href {\doibase 10.1007/JHEP11(2016)174} {\bibfield  {journal}
  {\bibinfo  {journal} {JHEP}\ }\textbf {\bibinfo {volume} {11}},\ \bibinfo
  {pages} {174} (\bibinfo {year} {2016})},\ \Eprint
  {http://arxiv.org/abs/1512.07561} {arXiv:1512.07561 [hep-ph]} \BibitemShut
  {NoStop}%
\bibitem [{\citenamefont {Mehtar-Tani}\ \emph
  {et~al.}(2012{\natexlab{b}})\citenamefont {Mehtar-Tani}, \citenamefont
  {Salgado},\ and\ \citenamefont {Tywoniuk}}]{MehtarTani:2012cy}%
  \BibitemOpen
  \bibfield  {author} {\bibinfo {author} {\bibfnamefont {Y.}~\bibnamefont
  {Mehtar-Tani}}, \bibinfo {author} {\bibfnamefont {C.~A.}\ \bibnamefont
  {Salgado}}, \ and\ \bibinfo {author} {\bibfnamefont {K.}~\bibnamefont
  {Tywoniuk}},\ }\href {\doibase 10.1007/JHEP10(2012)197} {\bibfield  {journal}
  {\bibinfo  {journal} {JHEP}\ }\textbf {\bibinfo {volume} {1210}},\ \bibinfo
  {pages} {197} (\bibinfo {year} {2012}{\natexlab{b}})}\BibitemShut {NoStop}%
\bibitem [{\citenamefont {Milhano}\ and\ \citenamefont
  {Zapp}(2016)}]{Milhano:2015mng}%
  \BibitemOpen
  \bibfield  {author} {\bibinfo {author} {\bibfnamefont {J.~G.}\ \bibnamefont
  {Milhano}}\ and\ \bibinfo {author} {\bibfnamefont {K.~C.}\ \bibnamefont
  {Zapp}},\ }\href {\doibase 10.1140/epjc/s10052-016-4130-9} {\bibfield
  {journal} {\bibinfo  {journal} {Eur. Phys. J.}\ }\textbf {\bibinfo {volume}
  {C76}},\ \bibinfo {pages} {288} (\bibinfo {year} {2016})},\ \Eprint
  {http://arxiv.org/abs/1512.08107} {arXiv:1512.08107 [hep-ph]} \BibitemShut
  {NoStop}%
\bibitem [{\citenamefont {Casalderrey-Solana}\ \emph
  {et~al.}(2017)\citenamefont {Casalderrey-Solana}, \citenamefont {Gulhan},
  \citenamefont {Milhano}, \citenamefont {Pablos},\ and\ \citenamefont
  {Rajagopal}}]{Casalderrey-Solana:2016jvj}%
  \BibitemOpen
  \bibfield  {author} {\bibinfo {author} {\bibfnamefont {J.}~\bibnamefont
  {Casalderrey-Solana}}, \bibinfo {author} {\bibfnamefont {D.}~\bibnamefont
  {Gulhan}}, \bibinfo {author} {\bibfnamefont {G.}~\bibnamefont {Milhano}},
  \bibinfo {author} {\bibfnamefont {D.}~\bibnamefont {Pablos}}, \ and\ \bibinfo
  {author} {\bibfnamefont {K.}~\bibnamefont {Rajagopal}},\ }\href {\doibase
  10.1007/JHEP03(2017)135} {\bibfield  {journal} {\bibinfo  {journal} {JHEP}\
  }\textbf {\bibinfo {volume} {03}},\ \bibinfo {pages} {135} (\bibinfo {year}
  {2017})},\ \Eprint {http://arxiv.org/abs/1609.05842} {arXiv:1609.05842
  [hep-ph]} \BibitemShut {NoStop}%
\end{thebibliography}%

\end{document}